\documentstyle{article}

\begin{document}

\title{\normalsize{\bf{	SUDDEN TO ADIABATIC TRANSITION IN BETA DECAY}}}
\author{J. Chizma \, G. Karl \ and V. Novikov*\\ \ \small{\it{Department of Physics, University of Guelph, Guelph, CANADA, N1G 2W1}}\\ \small{(* on leave from ITEP, Moscow, RUSSIA)}}
\date{\small{May 1 1998}\\ [.5cm]  \parbox{5.5in}{\hspace{1em}\small{\indent We discuss effects in beta decays at very low beta energies, of the order of the kinetic energies of atomic electrons. As the beta energy is lowered the atomic response changes from sudden to adiabatic. As a consequence, the beta decay rate increases slightly and the ejection of atomic electrons (shake off) and subsequent production of X rays is turned off. We estimate the transition energy, and the change in decay rate. The rate increase is largest in heavy atoms, which have a small Q value in their decay. The X ray switch-off is independent of Q value.\\ P.A.C.S.:23.40.-s, 23.90.+w}}}
\maketitle
\vspace{.25in}
\begin{quote}
\underline{1.Introduction}
\end{quote}

\normalsize
Atomic electrons affect beta decays, giving small corrections to spectral shapes, decay rates, and ionization processes. This subject is over fifty years old, going back to screening factors~\cite{rose} and the sudden approximation ~\cite{migdal}. It is beyond the scope of this note to review all previous work. Recent theoretical work was stimulated by accurate experiments in tritium decay and in superallowed decays~\cite{durand,drukarev,brown}.

We discuss effects in the low energy region of the beta spectrum of heavy atoms (\(Z>>1\)), where the beta energy is of the order of atomic energies. In this energy region the atomic dynamics suffers a transition as the beta energy is reduced. We find, as the beta energy diminishes, a small increase in the beta decay rate and a concomitant decrease in the atomic excitation and ionization processes. The interaction of the beta particle with the atomic system changes from sudden to adiabatic. Although these effects are potentially present in all previous theoretical work, we are unaware of any explicit estimates in the literature. These contributions are neglected on account of their small size. These effects are indeed small, but still observable. We focus on processes with heavy atoms, where the change from Z to \(Z \pm 1\) is so small that perturbation theory is a good approximation.

Consider the decay of a nucleus with a single atomic electron, in the ground electronic state. After the decay, the daughter system may be in the ground state, or in excited states. The probabilities of various excited states are determined in the 'sudden' approximation ~\cite{migdal}, which is very accurate, except for low beta velocities when the probabilities of excitation diminish, and the daughter system is left in its ground state. Therefore at sufficiently low beta energies the energy release to the lepton system (atomic Q value) increases slightly. This energy increase is carried (at fixed beta energy) by the outgoing neutrino (or antineutrino), which therefore has higher energy than estimated in the sudden approximation. The higher neutrino energy is manifested through an increase in rate of decay, which should be observable. It is this slight increase in rate of decay which concerns us. For tritium the effect occurs at too low  beta energies to be observed easily, but for heavier (large Z) nuclei the effect should be observable.

There are two main points to consider. First, one has to estimate the extra amount of energy (\(\Delta E\)) gained by the neutrino when the atomic excitation switches from sudden to adiabatic. Second, one has to estimate the beta energy at which the switching occurs. We find (on the first point) that for every filled electronic shell in the atom undergoing beta decay there is one atomic unit of energy (\(\Delta E\)) to be gained by the neutrino at low beta energies. As for the switching energy, we find a typical beta energy of the same order of magnitude as the kinetic energy of the electron in the shell under consideration. This means that inner shells of heavy atoms will show the rate increase at a higher beta energy than light atoms or outer shells of heavy atoms. The relative increase in differential rate R, (at a given beta energy \(E_{\beta}\)) is 
\begin{equation}
\frac{\delta R}{R} = 2 \frac{\Delta E}{Q-E_{\beta}} \simeq  2 \frac{\Delta E}{Q}
\end{equation}
Therefore this effect is more important in heavy nuclei (larger \(\Delta E\)) which beta decay with small Q values. For example, with a Q value of 100 keV, and 4 filled shells contributing, we obtain a rate change of about \(2 \times 10^{-3}\) at low energies. The change in the total rate (life-time) is too small to be observable.
\begin{quote}
We sketch details below.
\end{quote}

\begin{quote}
\underline{2. Energy release \(\Delta E\) when changing from sudden to adiabatic regime}
\end{quote}

We start with an atom having a single electron in the 1s state. In the sudden regime the atomic wavefunction does not have time to change, so that the mean energy of the atomic system (\(\overline{E}_{sud}\)) after decay is, in atomic units:

\begin{eqnarray}
\overline{E}_{sud} & = & \langle \psi (Z)| H(Z+1)|\psi (Z) \rangle \nonumber \\
  & = & \langle \psi (Z)| H(Z) - \frac{1}{r} |\psi (Z) \rangle = \frac{- Z^2}{2} - \frac{1}{Z} \cdot \frac{2Z^2}{2} \\
 & = & \frac{-Z^2}{2} - Z \
\; \;\mbox{a.u. } \nonumber
\end{eqnarray}
while in the adiabatic regime the electron remains in the 1s state about the new nucleus.
\begin{equation}
\overline{E}_{ad} = - \frac{1}{2} (Z+1)^2 \
 \; \mbox{a.u.}
\end{equation}
therefore,
\begin{equation}
\Delta E = \overline{E}_{sud} - \overline{E}_{ad} = \frac{1}{2}\
\; \mbox{a.u.}
\end{equation}

Thus \(\Delta E\) is independent of Z and is \(\frac{1}{2}\) 
a.u. per electron. For two electrons in the shell the contribution of each electron is \(\frac{1}{2}\) a.u., as may be estimated in the independent particle approximation, using the same procedure as above, but with an effective \(Z^{\prime}\) for the electron wavefunction of the two particle system.

Therefore one obtains for the K shell which contain two electrons \(\Delta E\) = one atomic unit, about 27 eV. For higher shells the contribution per electron diminishes in the same proportion as the number of electrons per shell \((2n^2)\) so that the total energy change per atomic shell remains at one atomic unit, independent of nuclear charge Z.

\begin{quote}

\underline{3. Switch-over energy from sudden to adiabatic regime}

\end{quote}

Corrections to the sudden approximation and their effect on the beta decay of tritium have been discussed in the literature by Durand and Lopez~\cite{durand}, Drukarev and Strikman ~\cite{drukarev}, and by Brown and Zhai ~\cite{brown}, who were interested in the precise shape of the beta spectrum of tritium and superallowed decays. The tritium decay is important for determination of the neutrino mass. We are interested in this note about analogous issues in the decay of \underline{heavy atoms}, with large nuclear charge Z, where distortions in the beta spectrum are produced at low beta energies. 

We have looked at a number of models to describe the dynamics of the decay, and find the details model dependent, but the range of energy where the sudden regime changes is quite stable. We shall only give an estimate of this energy range, by computing perturbatively the corrections away from the wo extreme limits, sudden and adiabatic. We consider a nucleus of arbitrary Z (\(Z>>1\)), with a single atomic electron. The change in the Hamiltonian from Z to \(Z\pm 1\) is small, so that time dependence may be treated perturbatively (see also ref. 6). We shall assume that initially the electron is in the 1s state. In the adiabatic limit (small beta velocity v) the final atomic electron remains in the same state, so the probability \(W_1\) is unity. In the sudden approximation, this probability is the square of the overlap integral of the two wavefunctions ~\cite{migdal}:
\begin{equation} 
{W_1}^{sudd} = |\langle {\psi} _{1s}^{Z+1}| {\psi} _{1s}^{Z} \rangle |^2 \simeq 1 - \frac{3}{4} \frac{1}{Z^2} + \ldots
\end{equation}

We want to determine corrections to these two limits, as a function of beta velocity.We omit details, but give the main steps and results. We estimate the leading order changes in the sudden and adiabatic probabilities to be:
\setcounter{equation}{0}
\renewcommand{\theequation}{6\alph{equation}}
\begin{equation}
W_{1}^{Ad} = 1 - \frac{1}{Z^2} \left(\frac{3.45v}{v_{atomic}} \right)^4 + \ldots
\end{equation}
\begin{equation}
W_{1}^{Sud} = 1 - \frac{3}{4Z^2} + \frac{1}{Z^2} \left( \frac{v_{atomic}}{3.09v} \right) ^2 + \ldots
\end{equation}
\setcounter{equation}{6}
\renewcommand{\theequation}{\arabic{equation}} 
\hspace{-.75em}where \(v\) is the beta velocity and \(v_{atomic}\) is the electron velocity in the 1s shell of the atom with nuclear charge Z. These two expansions (in \(v\) and \(v^{-1}\)) approach each other at \(\overline{v}\simeq 0.3 v_{at}\) which we take as the switchover velocity from the sudden to the adiabatic regime.

We sketch the derivation of the expansions in the two variables \(v\) and \(v^{-1}\). In both cases one can start from the time dependent Schr\"{o}dinger equation in an 'instantaneous' basis~\cite{schiff};
\begin{equation}
\dot{a} _{k} (t) = \sum _{n\neq k} \frac{a_{n}(t)}{\omega_{kn}(t)} \left \langle \phi _k (t) | \frac{\partial H}{\partial t} |\phi _{n} (t) \right \rangle e^{i {\int} _0 ^ t d\tau \,{\omega}_{kn} (\tau)}
\end{equation}
where the wavefunctions \(\phi_{n} (t)\) are the instantaneous eigenfunctions of H(t), with eigenvalues \(\epsilon_{n} (t)\) and \(\omega_{kn} (t) = \epsilon_{k} (t) - \epsilon_{n} (t) \). Since the change in the Hamiltonian is small ( \( Z \rightarrow Z+1, \Delta H/ H \sim 1/Z\) ) one can however use wavefunctions \(\phi_{k}\), and energies which are time independent, and integrate the equations under the assumption that \(a_{n} (t) = \delta_{n1} \) at all times. Then one obtains for the final excitation amplitudes
\begin{equation}
a_{k} (\infty) =\frac{1}{\omega_{k1}} \int _0 ^ \infty dt \, e^{i \omega_{k1} t} \left \langle \phi_{k} | \frac{\partial H}{\partial t} | \phi_{1} \right \rangle
\end{equation}
 and the probability of no excitation is obtained from unitarity
\begin{equation}
W_{1} = |a_{1} (\infty)|^2 = 1 - \sum_{k\geq 2} |a_{k} (\infty)|^2
\end{equation}
This equation gives the formal solution to the probability \(W_{1}\) for both the sudden and adiabatic cases. As a model for the perturbation of the atom we take the potential due to an outgoing s-wave beta particle:
\begin{equation}
V(r,t) = \left\{\begin{array}{ll}
   \frac{e^2}{r} & \mbox{for \(r>vt\)} \\ \nonumber
   \\ \nonumber
   \frac{e^2}{vt} & \mbox{for \( r<vt \)}
\end{array}
\right.
\end{equation}
and therefore the operator \( \frac{\partial H}{\partial t} = - \frac{e^2}{vt^2} \) for \( 0<r<vt \), and zero otherwise. This gives
\begin{equation}
\begin{array}{ll}
\left \langle k|\frac{\partial H}{\partial t} | 1 \right \rangle & = \int _0 ^{vt} dr\, r^2 R_{k0}(r) R_{10}(r) \left(\frac{-e^2}{vt^2} \right) \\
  \\  
 & = - \frac{e^2}{vt^2} \frac{1}{3} (vt)^3R_{10} (0) R_{k0} (0) + O(v^4)
\end{array}
\end{equation}
and therefore, in the adiabatic case
\begin{equation}
a_{k}(\infty) = \frac{e^2v^2}{3\omega_{k1}^3} R_{10}(0) R_{k0}(0)
\end{equation}
and
\begin{equation}
|a_{1}|^2 = 1 - \sum_{2}^{\infty} \frac{e^4 v^4}{9 \omega_{k1}^{6}} |R_{10}(0) R_{k0}(0)|^2
\end{equation}
The total probability to excite bound states with \(k>1\) is thus
\begin{equation}
\sum _{k>2} |a_{k}(\infty)|^2 = \frac{2^{10}}{9} \frac{1}{Z^2} \left( \frac{v}{v_{at}} \right) ^4 \cdot 0.8288 \simeq \frac {1}{Z^2} \left( \frac{3.116v}{v_{at}} \right) ^4 + O(v^6)
\end{equation}
where we denote \(Ze^2 \) by \(v_{atomic}\). The continuum contribution is one half of this probability, so that we obtain eqn.(6a). 

To find corrections to the sudden limit, we can start also from formula (8), which at short times may be evaluated most conviently by changing the order of integrations and the limits of integration, as follows:
\begin{equation}
\begin{array}{ccc}
a_{k}(\infty)  =  \frac{1}{\omega_{1k}}\int_0^\infty dt \, e^{i\omega_{1k}t} \int_0^{vt} dr \, r^2 R_{10}(r)R_{k0}(r) \left( \frac{-e^2}{vt^2} \right) \nonumber \\
\\
   =  \frac{-e^2}{v\omega_{1k}} \int_0^\infty dr \, r^2 R_{10}(r)R_{k0}(r) \int_0^{v/r} dt \, \left( \frac{1}{t^2} \right) \left( 1+i\omega_{1k}t - \frac{1}{2}{\omega_{1k}}^2t^2 \ldots \right) \\
\\
  =  -\frac{1}{\omega_{1k}}\langle 1 |\frac{e^2}{r}|k \rangle + \frac{ie^2}{v} \langle 1|ln{r}|k \rangle - \frac{e^2\omega_{1k}}{2v^2}\langle 1|r|k \rangle + \ldots
 \nonumber
\end{array}
\end{equation}

The singularity at the lower limit of the second integral is only apparent. If we introduce a cut off \(\epsilon\), the first integral vanishes by orthogonality. The first term in the final result is the first order perturbative estimate, which at large Z is identical to the sudden approximation (see eg. ref 6). The next terms are the corrections to the sudden approximation, and are sufficient to obtain the result (6b) by using the unitarity equation (9).

We conclude that the switching from the sudden regime to the adiabatic regime occurs for \(W_{1}\) at a beta energy \(E_{trans}\) approximately
\begin{center}
\(E_{tr}\sim 0.1 \times 13.6 \hspace{0.5em}eV \times Z^2 \sim 1.4\hspace{0.25em} Z^2 \hspace{0.5em} eV\)
\end{center}
which for \(Z\sim 50\) would be in the region of 3 keV, and therefore be amenable to observation.

\begin{quote}

\underline{4. Other effects and final comments}

\end{quote}

There are two related effects. The first concerns the fate of the atomic electron which is being excited in the sudden regime. While the probability of excitation at large Z is small (0.75 \(Z^2\)), a significant fraction of this excitation is in the continuum, leading to  so called shake-off electrons. In the sudden regime, approximately half of the excitation is to the continuum, as the beta particle energy is lowered, this probability of shake-off is eventually turned-off. This could be monitored experimentally by measuring the correlation between the yield of (subsequent) atomic-X rays of the daughter atom with the primary beta energy. At low beta energies these atomic-X rays should slowly disappear. This should be observable experimentally, with the shake-off in the most strongly bound shell (K-shell) disappearing first, at the largest beta energies, followed later, at lower beta energies by L-shell X rays, etc. Although attempts at measuring the beta energy dependence of X ray yield are recorded in the literature, the experiments we are aware of ~\cite{isozumi} did not go to low enough energies to see this effect. We believe the effect should be observable with due diligence, though the yields of X rays are small.

The second effect is a very small change in the overall lifetime for the parent decay, but this should be too small to be observable. Even for a small Q value, Q\(\simeq\) 100 keV, the change in lifetime is of order \(10^{-10}\).

Finally we comment on the main shortcoming of our estimate. We have neglected exchange effects between the beta electron and the atomic electrons which should be especially important for the details of the behaviour in the energy region we consider. But the expansions away from the transition region, and the energy region where the transion occurs should be reliable.

\begin{center}

AKNOWLEDGMENTS

\end{center}

We are indebted for conversations and help to J.L. Campbell, N. Isgur, B.G. Nickel, J.J. Simpson and L. Stodolsky. One of us (VAN) wishes to thank the Aspen Centre for Physics and the Max Plank Institute, Munich, for their hospitality in 1997 where part of this work was carried out.

\end{document}